\documentclass[showpacs,preprintnumbers,amsmath,amssymb,aps,eqsecnum]{revtex4}
\usepackage[dvips]{graphics,color}
\usepackage{latexsym}
\usepackage{graphicx}
\def\be{\begin{equation}}
\def\ee{\end{equation}}
\def\bea{\begin{eqnarray}}
\def\eea{\end{eqnarray}}
\def\r{{\hat{\rho}}}
\def\L{{\cal{L}}}

\def\V{{\cal{V}}}

\begin{document}

\title{Coupled boundary and bulk fields in anti-de Sitter}

\author{Kazuya Koyama, Andrew Mennim and David Wands\\~}

\affiliation{Institute of Cosmology and Gravitation, University of
  Portsmouth, Portsmouth~PO1~2EG, United Kingdom}

\date{25 April 2005}

\begin{abstract}
  We investigate the dynamics of a boundary field coupled to a bulk
  field with a linear coupling in an anti-de Sitter bulk
  spacetime bounded by a Minkowski (Randall-Sundrum) brane.  An
  instability criterion for the coupled boundary and bulk system is
  found. There exists a tachyonic bound state when the coupling is
  above a critical value, determined by the masses of the brane and bulk
  fields and AdS curvature scale.
  This bound state is normalizable and localised near
  the brane, and leads to a tachonic instability of the system on large
  scales. Below the critical coupling, there is no tachyonic state and
  no bound state. Instead, we find quasi-normal modes which describe
  stable oscillations, but with a finite decay time. Only if the
  coupling is tuned to the critical value does
  there exist a massless stable bound state, as in the case of zero
  coupling for massless fields. We discuss the relation to
  gravitational perturbations in the Randall-Sundrum brane-world.
\end{abstract}

\pacs{04.50.+h, 11.10.Kk, 11.10.St \hfill hep-th/0504201}

\maketitle

\section{Introduction}

There has been huge interest over recent years in
higher-dimensional models involving degrees of freedom which
propagate on lower dimensional surfaces (or branes) in a
higher-dimensional spacetime (or bulk). In particular for
codimension-one branes in an anti-de Sitter (AdS) spacetime, bulk
fields have a zero-mode localised near the brane, and hence can
yield a lower-dimensional effective theory at low energies without
a compact extra dimension~\cite{RS99}.
In traditional Kaluza-Klein compactification the low energy
effective theory arises from the lowest eigenmodes of the discrete
spectrum of bulk fields. By contrast the lower-dimensional field
living on the boundary of a bulk spacetime can in principle couple
to an infinite tower of bulk states even at linear order in
perturbation theory~\cite{Reviews2}.

In practice the more massive bulk modes may be weakly coupled to
the boundary at low energies and hence the tower may be neglected.
But if we want to determine the quantum vacuum state of the
boundary field at all energies we need to be able to deal with
coupling to bulk modes. This is a practical problem in
cosmological models of inflation driven by an inflaton field on a
brane \cite{MWBH99}. The large-scale structure of our universe is
supposed to originate from small-scale vacuum fluctuations of the
inflaton field during inflation. But these inflaton fluctuations
on the brane start on arbitrarily small scales where they are
strongly coupled to the infinite tower of bulk graviton modes \cite{KLMW}. 
Therefore, a truncated tower may not give a useful description of the vacuum
state on small-scales from which the large-scale
structure of our universe is supposed to be derived.

To derive the vacuum state of a coupled boundary-bulk system we
must include the full spectrum of normalisable modes of the
system. We note that the spectrum of normalisable modes can
include discrete bulk eigenmodes with complex eigenvalues in
addition to the familiar modes with real eigenvalues that form a
complete orthonormal set of basis functions for any classical
solution.
Dubovsky et al.~\cite{Rubakov} studied quasi-local states with complex
eigenvalues for massive bulk fields in the anti-de Sitter
bulk bounded by a vacuum Randall-Sundrum brane-world.
This work has been extended to include the curvature of the brane~\cite{Himemoto},
and the self-coupling of the bulk field on the brane~\cite{LS}.
Recently, Seahra~\cite{Seahra} has identified the
quasi-normal modes of gravitational waves by seeking
complex eigenvalues of the bulk mode equation.
As far as we are aware there has be no attempt previously to
identify the bound states or quasi-normal modes of coupled
boundary-bulk field theories in an anti-de Sitter spacetime.
(See Ref.~\cite{Bucher} for a different approach.)

Cartier and Durrer~\cite{CD04} did consider the general solution for
gravitational wave modes coupled to an anisotropic stress on the
brane in a Randall-Sundrum model. They found that the system could
describe normalisable modes with the anisotropic stress growing
exponentially with respect to time on the brane. They went on to
suggest that instabilities might be generic in any orbifold
model for the bulk. But the generality of their argument is open
to doubt without including the self-consistent evolution of a
reasonable matter source on the brane, and without checking that
the instability is not driven by energy flux from the past Cauchy
horizon.

As a first attempt to study the dynamics of a boundary field
coupled to bulk fields in AdS we study a simple model of massive
bulk and boundary fields with a linear interaction term on the
brane defined in Section~\ref{sect:model}.
We begin by reviewing in Section~\ref{sect:Mink} the recent work of
George \cite{George} who considered an oscillator on the Minkowski
boundary coupled to a massive field in a Minkowski bulk. 
We then go on to extend this to a massive field on a Minkowksi brane
in an AdS bulk in Section~\ref{sect:AdS}, identifying resonant modes
corresponding to bound states and quasi-normal modes, and present a condition for the existence of tachyonic instabilities.
We explore analytic and numerical solutions for these resonant modes
in limiting cases in Section~\ref{sect:cases}. 
We summarise our results in Section~\ref{sect:conc} and discuss
possible implications for gravitational perturbations in
Randall-Sundrum brane-worlds.


\section{The model}
\label{sect:model}

In the $(d+1)$-dimensional bulk spacetime $\V$, with metric
$G_{MN}$, we consider a free scalar field $\phi$, with Lagrangian
\begin{equation}
\L_\phi = - \frac12 G^{MN} \phi_{,M}\phi_{,N} - \frac12 m^2 \phi^2
\,,
\end{equation}
and on the $d$-dimensional boundary $\partial\V$, with metric
$g_{\mu\nu}$, we consider a field $q$ with Lagrangian
\begin{equation}
\L_q = - \frac12 g^{\mu\nu} q_{,\mu}q_{,\nu} - \frac12 \mu^2 q^2
\,.
\end{equation}
For simplicity we consider a linear interaction between the boundary
field and the bulk field on the boundary
\begin{equation}
\L_{\rm int} = - \beta \phi q \,.
\end{equation}
Note that the system is symmetric under $\beta\to-\beta$ and
$\phi\to-\phi$ or $q\to-q$. Henceforth we take $\beta\geq0$ without
loss of generality.
The total action is thus
\begin{equation}
 S = \int_\V d^{d+1}x \sqrt{-G} \L_\phi + \int_{\partial\V}
  d^dx \sqrt{-g} \left[ \L_q + \L_{\rm int} \right]
 \,.
\end{equation}

The coupled wave equations are then
\begin{equation}
\label{eomq}
^{(d)}\Box q = \mu^2 q + \beta \phi_0 \,,
\end{equation}
on the brane, where $\phi_0$ denotes the value of $\phi$ at the boundary,
and a free wave equation for $\phi$ in the bulk
\begin{equation}
\label{eomphi}
^{(d+1)}\Box \phi = m^2 \phi \,,
\end{equation}
subject to the boundary condition \footnote{Note that we differ
from Ref.~\cite{George} in two respects: firstly Eq.~(\ref{bc})
has a factor of 2 difference because we are assuming $Z_2$
symmetry about the brane, so that $[\phi_0']_-^+=2\phi_0$;
secondly we take $y>0$ in the bulk rather than $y<0$.}
 \begin{equation}
 \label{bc}
 \phi_0' =  \frac{\beta}{2} q \,,
 \end{equation}
at the brane, where $\phi_0'$ signifies the normal derivative at
the boundary.

Note that if we were considering a linear interaction between two
massive boundary fields with total Lagrangian
 \begin{equation}
\label{L2qs}
 \L = \L_{q_1} + \L_{q_2} - \bar\beta q_1 q_2 \,,
 \end{equation}
then we could easily diagonalise the mass-matrix and identify the
effective squared-masses
 \begin{equation}
 m_\pm^2 =
 \frac{m_1^2+m_2^2\pm\sqrt{(m_1^2+m_2^2)^2+4\bar\beta^2-4m_1^2m_2^2}}{2}
 \,,
 \end{equation}
on of which becomes negative for
 \begin{equation}
\label{critbarb}
 \bar\beta^2 > m_1^2 m_2^2 \,,
 \end{equation}
signalling a tachyonic instability.
Thus, we should not be surprised to find an instability for large
coupling.

\section{Minkowski bulk}
\label{sect:Mink}

George \cite{George} recently considered an oscillator on a flat
boundary coupled to a field in a flat two-dimensional Minkowski spacetime.
We write the corresponding bulk metric in $(d+1)$-dimensional
Minkowski spacetime as
 \begin{equation}
ds^2 = -dt^2 + \delta^{ij} dx_i dx_j + dy^2 \,.
 \end{equation}
The coupled wave equations (\ref{eomq}) and (\ref{eomphi}) thus
become
 \begin{eqnarray}
  \label{eomqMink}
\ddot{q} + k^2 q &=& -\mu^2 q - \beta \phi_0 \,,\\
 \label{eomphiMink}
\ddot\phi + k^2 \phi &=& \phi'' - m^2\phi \,,
 \end{eqnarray}
where a prime denotes a derivative with respect to $y$ and a dot with respect to $t$,
subject to the boundary condition (\ref{bc}), and we consider a single
Fourier mode with wavenumber $k$ on the $(d-1)$-dimensional
sub-space.

\subsection{General solution}

Formally the general solution can be written as a sum over modes \cite{George}
 \begin{eqnarray}
\label{Minksoln}
q(t) &=& \int_{-\infty}^\infty d\rho \, C_\rho  T_\rho(t) \,, \nonumber \\
 \phi(t,y) &=& \int_{-\infty}^\infty d\rho \,A_\rho \,e^{i\rho y} \,T_\rho(t) \,.
 \end{eqnarray}
The bulk wave equation (\ref{eomphiMink}) requires
 \begin{equation}
\label{ddotT}
\ddot{T}_\rho = -\omega_\rho^2 T_\rho \,,
 \end{equation}
where, $\omega_\rho$ is given by the dispersion relation
 \begin{equation}
\label{omegarho}
\omega_\rho^2 = \rho^2 + m^2 + k^2 \,.
 \end{equation}
In addition the boundary wave equation (\ref{eomqMink})
and the boundary condition for the bulk field (\ref{bc})
require that the following consistency relations between
the coefficients
\begin{eqnarray}
\label{CAMink}
&\big( \rho^2 + m^2 - \mu^2 \big) \big(C_\rho+C_{-\rho}\big)
= \beta \big(A_\rho+A_{-\rho}\big) \,,&\\
&i\rho\big(A_\rho-A_{-\rho}\big)=\frac{\beta}{2}\big(C_\rho+C_{-\rho}\big)\,.&
\label{CBMink}
\end{eqnarray}
Eliminating $(C_\rho+C_{-\rho})$ from these two constraints we obtain a
relation between the coefficients in the bulk mode
 \begin{equation}
\label{ABMink}
\left[i\rho\big(\rho^2+m^2-\mu^2\big)-\frac{\beta^2}{2}\right]A_\rho=
\left[i\rho\big(\rho^2+m^2-\mu^2\big)+\frac{\beta^2}{2}\right]A_{-\rho}\,.
\end{equation}
The constraints interpolate between Neumann boundary
conditions at weak coupling ($\beta\to0$) and Dirichlet boundary
conditions at strong coupling ($\beta^2\to\infty$) \cite{George}.

\subsection{Resonant modes}

Conventionally, $\rho$ is assumed to be real; however,
resonant modes with complex $\rho$ play a crucial role 
in a scattering process. Resonant modes can be found 
by demanding that the bulk field is purely out-going 
at infinity $y \to \infty$. We take the temporal mode 
function as 
\begin{equation}
T_{\rho}(t) = e^{ -i \omega_{\rho} t},
\end{equation}
and impose the condition $\phi(y,t) \to e^{i \rho y}
e^{ -i \omega_{\rho} t}$ as $y \to \infty$. In order to satisfy
this condition, we must have
\begin{equation}
A_{-\rho} =0.
\end{equation}
Substituting this constraint into equation (\ref{ABMink}) yields
\begin{equation}
\label{cubicrho}
\rho ( \rho^2 + m^2 - \mu^2 ) + i \frac{\beta^2}{2} =0 \,.
 \end{equation}
The out-going boundary condition are satisfied   when the real parts
of $\rho$ and $\omega_\rho$ have opposite signs, i.e., when
\begin{equation}
\Re (\rho) \Re (\omega_{\rho}) > 0\,.
\label{out-going}
\end{equation}

Resonant modes correspond to the poles of the Green function \cite{Rubakov}.
Let us compute the 5D Green function 
$\bigtriangleup(x^{\mu},y;x^{\mu'},y')$ that satisfies
\begin{equation}
\left( \partial_y^2 - p^2 \right) 
\bigtriangleup(\rho,y,y') = \delta(y-y')\,,
\end{equation}
where we performed a Fourier transformation along the brane
\begin{equation}
\bigtriangleup(x^{\mu},y;x^{\mu'},y') = \int \frac{d^{d-1} p}{(2
\pi)^{d-1}} e^{i p \cdot (x-x')} \bigtriangleup(\rho,y,y').
\end{equation}
and $\rho^2=-p^2=\omega_{\rho}^2-m^2-k^2$. The boundary condition for the
Green function at the brane $y=0$ should be determined by the
boundary conditions for the scalar field in question:
\begin{eqnarray}
\phi'(\rho) &=& \frac{\beta}{2}q(\rho), \nonumber\\
\phi(\rho) &=& \frac{1}{\beta} (\rho^2+m^2 -\mu^2) q(\rho),
\end{eqnarray}
where $\phi(\rho)$ and $q(\rho)$ are the Fourier modes of
$\phi(0,x^{\mu})$ and $q(x^{\mu})$ respectively. In order to
satisfy these boundary conditions, we must impose the boundary
condition for the Green function as
\begin{equation}
\left[ (\rho^2+m^2-\mu^2) \partial_y \bigtriangleup - \frac{\beta^2}{2}
\bigtriangleup \right ]_{y=0}=0.
\end{equation}
We also impose the out-going boundary condition at infinity.
The Green function has a simple form if one of the arguments is on
the brane;
\begin{equation}
\bigtriangleup(\rho,y,0) =
\frac{\rho^2+m^2 - \mu^2}{i \rho (\rho^2 + m^2 - \mu^2) - \frac{\beta^2}{2}}
 e^{i \rho y}.
\end{equation}
We notice that the condition for Eq.~(\ref{cubicrho}) is the condition
for the poles of the Green function.

The causal boundary condition for the Green function is determined
by the way in which the contour of integration is closed around the
poles. In order to impose the retarded boundary condition we
must close the integration contour in the upper half complex
$\omega_{\r}$ plane  for $t < t'$ and subtract the contributions
of the poles in the upper half complex $\omega_\r$ plane. For $t >
t'$, we can close the integration contour on the lower half of the
complex $\omega_{\r}$ plane and the late time behavior of the
scalar field can be understood by investigating the structure of
singularities such as poles in the integrand.

\begin{figure}[htbp]
\centerline{
\includegraphics[width=10cm]{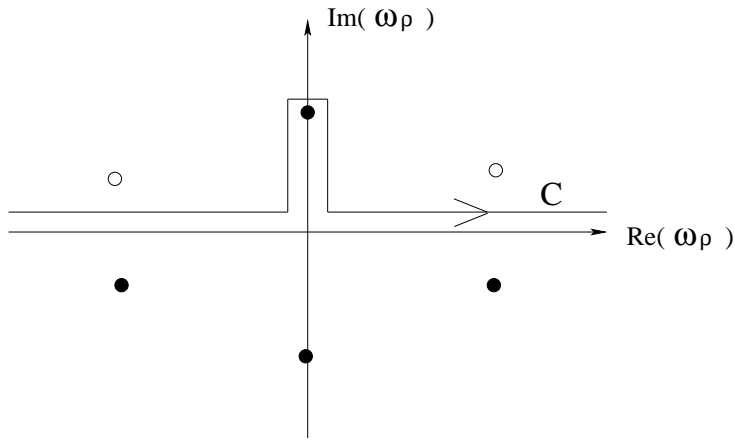}}
\caption{Location of poles in a complex $\omega_{\rho}$ plane
for $\mu > m$ and $\beta> 2m \mu^2$ with $k =0$.
Filled circle show poles with the out-going boundary conditon
and open circles show poles with the in-going boundary
condition. The contour is chosen so that the retarded boundary
condition is satisfied in the case of the out-going boundary condition.}
\label{MinkowskiPole}
\end{figure}

Fig.~\ref{MinkowskiPole} shows the solutions for $\omega_{\rho}$ in a complex
$\omega_{\rho}$ plane for $\mu > m$ and $\beta> 2m \mu^2$.
The integration contour must be taken like $C$ in order to impose
retarded boundary condition.
We find two kinds of poles described below.

\subsubsection{Bound state}
The bound state is a mode that is exponentially decaying and localised
near the brane.
Setting $\rho=i\sigma$ we see that we are seeking roots of the
cubic equation
 \begin{equation}
\label{cubic}
 \sigma^3 + \sigma ( \mu^2-m^2 ) -\frac{\beta^2}{2} = 0 \,.
 \end{equation}
with positive real part.

Equation (\ref{cubic}) always has one (and only one)
positive real root \cite{George}.
This will describe stable oscillations if $\omega_\rho^2>0$ which,
from Eq.~(\ref{omegarho}), requires $\sigma^2<m^2$ as $k\to0$, or equivalently
\cite{George}
 \begin{equation}
\beta^2 < 2 m \mu^2 \,.
 \end{equation}
On the other hand the bound state describes an unstable mode which
can grow exponentially in time above a critical coupling
$\beta^2_{\rm crit}=2m\mu^2$. In this case, we have two poles 
on the imaginary axis in the complex $\omega_{\rho}$ plane:
a pole in the upper half plane describes an unstable mode.

\subsubsection{Quasi-normal modes}
In addition to the bound state, there are solutions with complex
$\omega_{\rho}$ located in the lower half
$\omega_{\rho}$ plane. Thus, these solutions lead to an exponentially
decaying mode in time. We should note that this conclusion depends on
the boundary condition.  If we impose an in-going boundary condition
so that $\phi(y,t) \to e^{- i \rho y} e^{- i \omega_{\rho} t}$ with
the conditions Eq.~(\ref{out-going}), poles are located at the
upper-half complex $\omega_{\rho}$ plane, leading to an exponentially
growing mode in time. However, a physical boundary condition, i.e., the
out-going boundary condition under which there is no energy flux from infinity,
can eliminate these unstable modes, therefore, these instabilities do not describe a
physical instability of the system.
On the other hand, an unstable bound state for $\beta >
\beta_{\rm crit}$ cannot be eliminated by a choice of the boundary
condition, indicating a physical instability of the system, i.e., a
tachyonic instability.

\section{Anti-de Sitter bulk}
\label{sect:AdS}

We now extend the analysis to consider a boundary field coupled to a
massive bulk field in an Anti-de Sitter (AdS) bulk spacetime, as in
the Randall-Sundrum brane-world \cite{RS99}.
We write the corresponding bulk metric, with Anti-de Sitter curvature
scale $\ell$, as
 \begin{equation}
\label{AdS} ds_{d+1}^2 = e^{-2|y|/\ell}\left( -dt^2 + \delta^{ij}
dx_i dx_j
 \right) + dy^2
 = \frac{\ell^2}{z^2} \left( -dt^2 + \delta^{ij} dx_i dx_j + dz^2
 \right) \,.
 \end{equation}
Note that the boundary is now located at $z=\ell$, and the surface
$\ell\leq z < +\infty$ at fixed time $t$ defines a Cauchy surface, with
$t\to-\infty$ the past Cauchy horizon~\footnote{We are only concerned with
the region of AdS up to the future Cauchy horizon, not the whole covering space.}.

The coupled wave equations (\ref{eomq}) and (\ref{eomphi}) become
 \begin{eqnarray}
 \label{eomqAdS}
\ddot{q} + k^2 q &=& -\mu^2 q - \beta \phi_0 \,,\\
 \label{eomphiAdS}
\ddot\phi + k^2 \phi &=& \phi'' - \frac{3}{z} \phi' -
\frac{m^2\ell^2}{z^2} \phi \,,
 \end{eqnarray}
subject to the same boundary condition (\ref{bc}) as before,
where a prime now denotes derivatives with respect to $z$,
and we again consider a single Fourier mode with wavenumber $k$ on the
$(d-1)$-dimensional sub-space.

\subsection{General solution}

Formally the general solution can be written as a sum over modes
 \begin{eqnarray}
 \label{AdSsoln}
q(t) &=& \int_{0}^\infty d\r \, C_\r T_\r(t) \,, \nonumber \\
\phi(t,z) &=& \left(\frac{z}{\ell}\right)^2
\int_{0}^\infty d\r \, \left[ 
A_\r \,H^{(1)}_\nu (\r z) +B_\r  \, H^{(2)}_{\nu} (\r z) 
\right ] \, T_\r(t) \,.
 \end{eqnarray}
where $H_\nu^{(1)}$ and $H_\nu^{(2)}$ are Hankel functions of first and second kind
and order $\nu=\sqrt{4+m^2\ell^2}$.
The bulk wave equation (\ref{eomphiAdS}) then yields the time
dependence for $T_\r(t)$
 \begin{equation}
\label{ddotTr} \ddot{T}_\r = -\omega_\r^2 T_\r \,,
 \end{equation}
where $\omega_\r^2$ is given by the dispersion relation
 \begin{equation}
\label{omegar} \omega_\r^2 = \r^2 + k^2 \,.
 \end{equation}
The boundary wave equation (\ref{eomqAdS}) and the boundary
condition for the bulk field at the brane (\ref{bc}) impose the
conditions
\bea
&\big(\r^2-\mu^2\big) C_\r  = \beta \left[
A_\r H^{(1)}_\nu(\r\ell)+B_{\r} H^{(2)}_\nu(\r\ell)\right] \,,&\\
&\frac{\beta}{2} C_\r=
A_\r \left[\frac{2-\nu}{\ell}H^{(1)}_\nu(\r\ell)+
\r  \,H^{(1)}_{\nu-1}(\r\ell)\right]+
B_{\r} \left[\frac{2-\nu}{\ell}H^{(2)}_\nu(\r\ell)+\r
\,H^{(2)}_{\nu-1}(\r\ell)\right]\,,&
\eea
from which we eliminate $C_\r$ to get
\begin{eqnarray}
\label{generalAB}
A_\r  \Big[\r (\r^2-\mu^2)\,
H^{(1)}_{\nu-1}(\r\ell)&+&\Big\{\frac{2-\nu}{\ell}\big(\r^2-\mu^2\big)
+ \frac{\beta^2}{2}\Big\}H^{(1)}_\nu(\r\ell)\Big] \nonumber\\
=-
B_{\r}\Big[\r(\r^2-\mu^2)\,
H^{(2)}_{\nu-1}(\r\ell)&+&\Big\{\frac{2-\nu}{\ell}\big(\r^2-\mu^2\big)
+ \frac{\beta^2}{2}\Big\}H^{(2)}_\nu(\r\ell)\Big]\,.
\end{eqnarray}
We have used various well-known relations for Bessel functions and their derivatives
which can be found in Ref.~\cite{GradRyz}.

\subsection{Resonant modes}
In order to find resonant modes, we impose that
the modes become plane waves at large $z$, 
\begin{equation}
\phi_\r(z) \to K_\r z^{3/2} e^{i\r z-i\theta} \,,
\end{equation}
where $\theta$ is a phase and $K_\r$ is a constant.
Then we must have $B_{\r}=0$ and Eq.(\ref{generalAB}) gives
\begin{equation}
\r(\r^2 -\mu^2) H^{(1)}_{\nu-1} (\r \ell) +
\left[\frac{2-\nu}{\ell}(\r^2-\mu^2)-\frac{\beta^2}{2}\right]
H^{(1)}_{\nu}(\r \ell) =0,
\label{consistency}
\end{equation}
with $\Re(\r) >0$. 
Note that if we had instead imposed the condition $\phi \propto
z^{3/2}e^{-i\r z+i\theta}$, we would have obtained the condition
\begin{equation}
\r(\r^2 -\mu^2) H^{(2)}_{\nu-1} (\r \ell) +
\left[ \frac{2-\nu}{\ell}(\r^2-\mu^2)-\frac{\beta^2}{2}\right]
H^{(2)}_{\nu}(\r \ell) =0 \,.
\label{consistency2}
\end{equation}
The solutions to which are just the complex conjugates of
the solutions to Eq.~(\ref{consistency}).

Again, solutions to Eq.~(\ref{consistency}) are related 
to poles of the Green function. 
Let us compute the 5D Green function
$\bigtriangleup(x^{\mu},z;x^{\mu'},z')$ that satisfies
\begin{equation}
\frac{z^2}{\ell^2} \left(\partial_z^2-\frac{3}{z} \partial_z - p^2
\right)\bigtriangleup(\r,z,z') = \left( \frac{z}{\ell} \right)^5
\delta(z-z').
\end{equation}
where we performed a Fourier transformation along the brane
\begin{equation}
\bigtriangleup(x^{\mu},z;x^{\mu'},z') = \int \frac{d^4 p}{(2
\pi)^4} e^{i p \cdot (x-x')} \bigtriangleup(\r,z,z').
\end{equation}
and $\r^2=-p^2=\omega_{\r}^2-k^2$ (compare to Ref.~\cite{Giddings}).
The boundary condition for the Green function at the brane $z=\ell$ is given by
\begin{equation}
\left[ (\r^2-\mu^2) \partial_z \bigtriangleup - \frac{\beta^2}{2}
\bigtriangleup \right ]_{z=\ell}=0.
\end{equation}
We also impose the boundary condition at the Cauchy horizon of the
AdS spacetime so that the waves are out-going. The
Green function has a simple form if one of the arguments is on the
brane;
\begin{equation}
\bigtriangleup(\r,z,\ell) = \left(\frac{z}{\ell} \right)^2
\frac{(\r^2-\mu^2) H^{(1)}_{\nu}(\r z)}
{\r (\r^2-\mu^2)\,
H^{(1)}_{\nu-1}(\r\ell)+\Big[ \frac{2-\nu}{\ell}\big(\r^2-\mu^2\big)
+ \frac{\beta^2}{2}\Big] H^{(1)}_\nu(\r\ell)}.
\end{equation}
As in the Minkowski case, the condition for the bound states is the condition
for the poles of the Green function 
\begin{equation}
\bigtriangleup(\r,\ell,\ell)^{-1}=0.
\label{consistencygreen}
\end{equation}

\subsubsection{Bound states}
The AdS curvature in the bulk dramatically changes the condition for
the existence of a bound state.
Setting $\rho=i\sigma$, we seek the solution for Eq.~(\ref{consistency}).
We can use the relation
\begin{equation}
\label{K}
K_\nu(x)=\frac{i\pi}{2}e^{i\pi\nu/2}H_\nu^{(1)}(ix)
\end{equation}
where $K_\nu(x)$ is a modified Bessel function~\cite{GradRyz},
to write Eq.~(\ref{consistencygreen}) as
\begin{equation}
\sigma \frac{K_{\nu-1} (\sigma\ell)}{K_{\nu} (\sigma \ell)} +
\frac{(\nu-2)}{\ell} -\frac{\beta^2}{2} \frac{1}{(\sigma^2 + \mu^2)}=0.
\end{equation}
The existence of a positive real root depends on the value of 
the coupling. 
There is always one (and only one) positive real root
for $\beta>\beta_{\rm crit}$, where
\begin{equation}
\beta_{\rm crit}^2 = 2(\nu-2) \mu^2 \ell^{-1} \,.
\label{crit}
\end{equation}
As in the Minkowski bulk case, this describes an unstable mode
with $\omega_{\r}^2 <0$ when $k=0$.
Note that $\beta_{\rm crit}$ becomes $\beta_{\rm crit} = 2 m^2 \mu$
for $m \ell \gg 1$, which agrees with the critical coupling in the
Minkowski case. For $\beta=\beta_{\rm crit}$, there is a massless
bound state $\sigma=0$.
Unlike the Minkowski bulk case, there
is no positive real root for a small coupling $\beta < \beta_{\rm crit}$;
instead we find quasi-normal modes with complex $\r$.

\subsubsection{Quasi-normal modes}
We have a rich structure of quasi-normal modes for an AdS bulk.
Indeed, the bulk scalar field has an infinite number of quasi-normal
modes even without a coupling to the brane field~\cite{Seahra}.  As was shown in
Ref.~\cite{Rubakov}, if we introduce a mass to bulk fields in AdS
spacetime, we find quasi-localized states with finite width.
Quasi-normal modes with complex eigenvalue $\r$ describe metastable
states that decay into continuum modes within a finite time.  With
zero coupling, $\beta=0$, the condition for quasi-normal modes for the
bulk scalar field is given by
\begin{equation}
\r \frac{H^{(1)}_{\nu-1} (\r \ell)}{ H^{(1)}_{\nu}(\r \ell)}  +
\frac{2-\nu}{\ell} =0.
\label{consistencyno}
\end{equation}
At low energies $m \ell \ll 1$, the solution for $\r$ is given by
\begin{equation}
\r^2=\frac{m^2}{2}+ i\Gamma, \quad \Gamma= -\frac{\pi}{16} m^2 (m
\ell)^2,
\label{bulkbound1}
\end{equation}
with $\Re(\r) >0$. 
If we take the temporal mode function to be $T_\r(t)=e^{-i \omega_{\r} t}$, 
we must impose $\Re(\omega_{\r}) >0$ to satisfy the out-going 
boundary condition. 
On the other hand, if $\Re(\omega_{\r}) <0$, we must impose
$\phi \propto z^{3/2} e^{-i \r z}$ with $\Re(\r) >0$.
A solution for $\r$ at low energies becomes
\begin{equation}
\r^2=\frac{m^2}{2}- i\Gamma, \quad \Gamma= -\frac{\pi}{16} m^2 (m
\ell)^2,
\label{bulkbound2}
\end{equation}
with $\Re(\r) <0$. 

\begin{figure}[htbp]
\centerline{
\includegraphics[width=10cm]{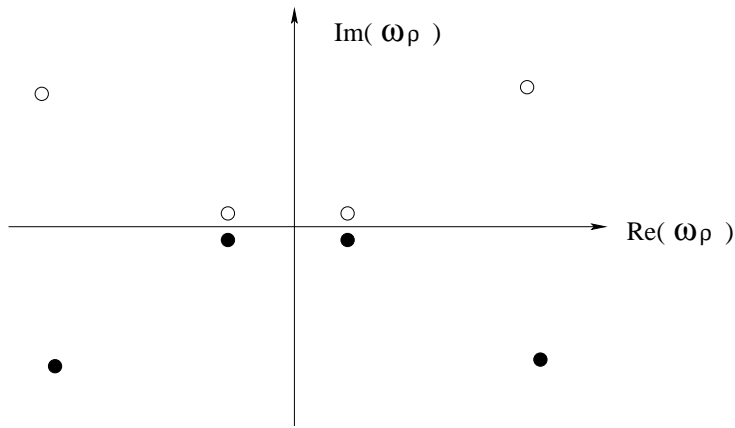}}
\caption{Location of poles in a complex $\omega_{\rho}$ plane 
for massive scalar fields with $k =0$. 
Filled circles show poles with the out-going boundary condiiton 
and open circles show poles with the in-going boundary 
condition. There are an infinite number of poles.}
\label{AdsPole}
\end{figure}

These poles are located in the lower half $\omega_{\rho}$ plane 
(see Fig.~\ref{AdsPole}) and these solutions lead to exponentially decaying
modes in time. Note that we could have exponentially growing 
modes in time if we imposed an in-going boundary condition, just as 
in the Minkowski bulk case. However, an in-going boundary condition 
implies energy flux from the past Cauchy horizon of the AdS spacetime.
Therefore, this is not a physical boundary condition.

A non-zero coupling to the brane field changes the location of the poles 
in a complicated way. We will study the effect of the coupling 
in the next section in various cases.

\section{Bound states and quasi-normal modes in limiting cases}
\label{sect:cases}

\subsection{High-energy limit}

Let us first consider the case where $m\ell \gg 1, \mu\ell \gg 1$
and search for solutions with $\r\ell \gg 1$. In this limit, we
can show
\begin{equation}
\frac{\r H_{\nu-1}^{(1)}(\r\ell)}{H_{\nu}^{(1)}(\r\ell)} 
+ \frac{2 - \nu}{\ell} \to i
\sqrt{\r^2 - m^2}.
\end{equation}
The condition for the bound state becomes the same as Minkowski
bulk case, equation (\ref{cubicrho}) if we make the identification
\begin{equation}
\rho^2 = \r^2 - m^2.
\end{equation}
Thus, we recover the Minkowski results in the limit when the AdS
length scale is much larger than the Compton wavelengths of both
the boundary and bulk fields.

\subsection{Low energy limit}

Let us now consider the low-energy limit where mass scales are
much less that the AdS scale ($m\ell \ll 1$ and $\mu\ell \ll 1$)
and search for bound states with $|\r\ell| \ll 1$. We can expand
the Hankel functions for small arguments and the condition for a
bound state (\ref{consistency}) becomes
\begin{equation}
(\r^2-\mu^2)\left(\r^2 +i \pi \frac{\r^4 \ell^2}{4} - \frac{m^2}{2}
\right)
 =\ell^{-1} \beta^2,
\end{equation}
where we assumed $m^2 \neq 0$. At weak coupling,
$\beta^2/\ell\to0$, there are two independent fields. One is the
brane field with $\r=\mu$ and the other is the quasi-bound state of the
bulk field with Eqs.~(\ref{bulkbound1}), (\ref{bulkbound2})

Neglecting the finite decay width, suppressed by $m\ell \ll 1$, we
can construct the effective coupled equations for the brane field
$q(t)$ and the bulk fields on the brane $\phi(0,t)$
 \begin{eqnarray}
\ddot{\phi}+k^2 \phi + \frac{1}{2}m^2 \phi &=& -\ell^{-1} \beta q, \nonumber\\
\ddot{q}+k^2 q +\mu^2 q &=& -\beta \phi.
 \end{eqnarray}
This $d$-dimensional effective theory coincides with the example of
two massive boundary fields with linear coupling defined in
Eq.~(\ref{L2qs}) if we identify $m_1^2 = \mu^2$, $m_2^2 = m^2/2$ and
$\bar{\beta}=\beta/\ell$. We would expect to be unstable above a
critical coupling given by Eq.~(\ref{critbarb}) which corresponds to
$\beta^2 > \ell m^2\mu^2/2$. Note that this critical coupling agrees
with Eq.~(\ref{crit}) for $m\ell \ll 1$. 

However, this effective theory neglects the slow decay indicated by
the imaginary part of the effective mass $\r$ suppressed by $\r l \ll
1$.  In the following sections, we study the effect of this small
imaginary part induced from the curvature of the AdS spacetime.

\subsection{Massless bulk and brane fields}

Let us consider the simple case where $m=\mu=0$. Henceforth, we will use units
where $\ell=1$, so that the consistency relation,
Eq.~(\ref{consistency}), for massless fields can be written as 
\begin{equation}
\r^3 H^{(1)}_{1} (\r)-\frac{\beta^2}{2} H^{(1)}_{2}(\r) = 0 \,.
\end{equation}
When $\beta=0$ there is no coupling and there is a zero mode solution
$\r=0$.  If the coupling is turned on, this will no longer necessarily
be a solution.  For small $\beta$ there should be a root near $\r=0$,
so we will investigate how the consistency relation is perturbed.  The
Hankel functions can be expanded for $|\r| \ll 1$ by
\begin{eqnarray}
H^{(1)}_2(\r)&=&-\frac{4i}{\pi\r^2}-\frac{i}{\pi}+
\frac{i}{4\pi}\r^2\log\left(\frac{\r}{2}\right)
+\frac{2\pi+4Ci-3i}{16\pi}\r^2+\cdots,\\
H^{(1)}_1(\r)&=&-\frac{2i}{\pi\r}+\frac{i}{\pi}\r\log\left(\frac{\r}{2}\right)
+\frac{\pi+2Ci-i}{2\pi}\r+\cdots,
\end{eqnarray}
where $C\approx0.577$ is the Euler constant~\cite{GradRyz}. 
So the consistency relation can be approximated as
\begin{equation}
\frac{2i}{\pi}\r^2-\frac{i}{\pi}\r^4\log\left(\frac{\r}{2}\right)
-\frac{\pi+2Ci-i}{2\pi}\r^4+\cdots=\frac{2i\beta^2}{\pi\r^2}+\frac{i\beta^2}{2\pi}
-\frac{i}{8\pi}\beta^2\r^2\log\left(\frac{\r}{2}\right)
+\frac{3i-4Ci-2\pi}{32\pi}\beta\r^2+\cdots\,.
\end{equation}
To lowest order in $\r$ this has solution
\begin{equation}
\r^4=\beta^2 \qquad \Rightarrow \qquad\r^2=\pm\beta
\qquad \Rightarrow \qquad \r=e^{i\pi n/4}\sqrt{\beta}, \quad n=0,1,2,3.
\end{equation}
if we continue the solution $\r=\sqrt\beta$ by adding higher order
terms in $\beta$ we get
\begin{equation}
\r=\sqrt\beta+\frac{1}{16}\beta^{3/2}\log\beta+
\frac{1}{8}\big(C-\log2\big)\beta^{3/2}-\frac{i\pi}{16}\beta^{3/2}+\cdots\,.
\end{equation}
Using the dispersion relation (\ref{omegar}) we see that the frequency
is not real, but is given approximately by
\begin{equation}
\omega=\pm\sqrt{k^2+\beta} - \frac{i \pi}{16} \frac{\beta^2}{\sqrt{k^2+\beta}}
\end{equation}
again, keeping only the lowest order real and imaginary parts.
Note that for $\Re(\omega_{\r}) <0$, we must use the solution for 
Eq.~(\ref{consistency2}) to impose the out-going boundary condition. 
We see that the poles are located at a lower-half plane and these
solutions are quasi-normal modes that describe massive metastable 
states.
For the root near $i\sqrt{\beta}$, we get
\begin{equation}
\r=i\sqrt\beta-i\frac{1}{16}\beta^{3/2}\log\beta-
i\frac{1}{8}\big(C-\log2\big)\beta^{3/2},
\end{equation}
where the real part vanishes. This is the massive bound state which
leads to a tachyonic instability.

\begin{figure}[htpb]
\begin{minipage}[t]{8cm}
\begin{center}
\resizebox{8cm}{8cm}{\includegraphics{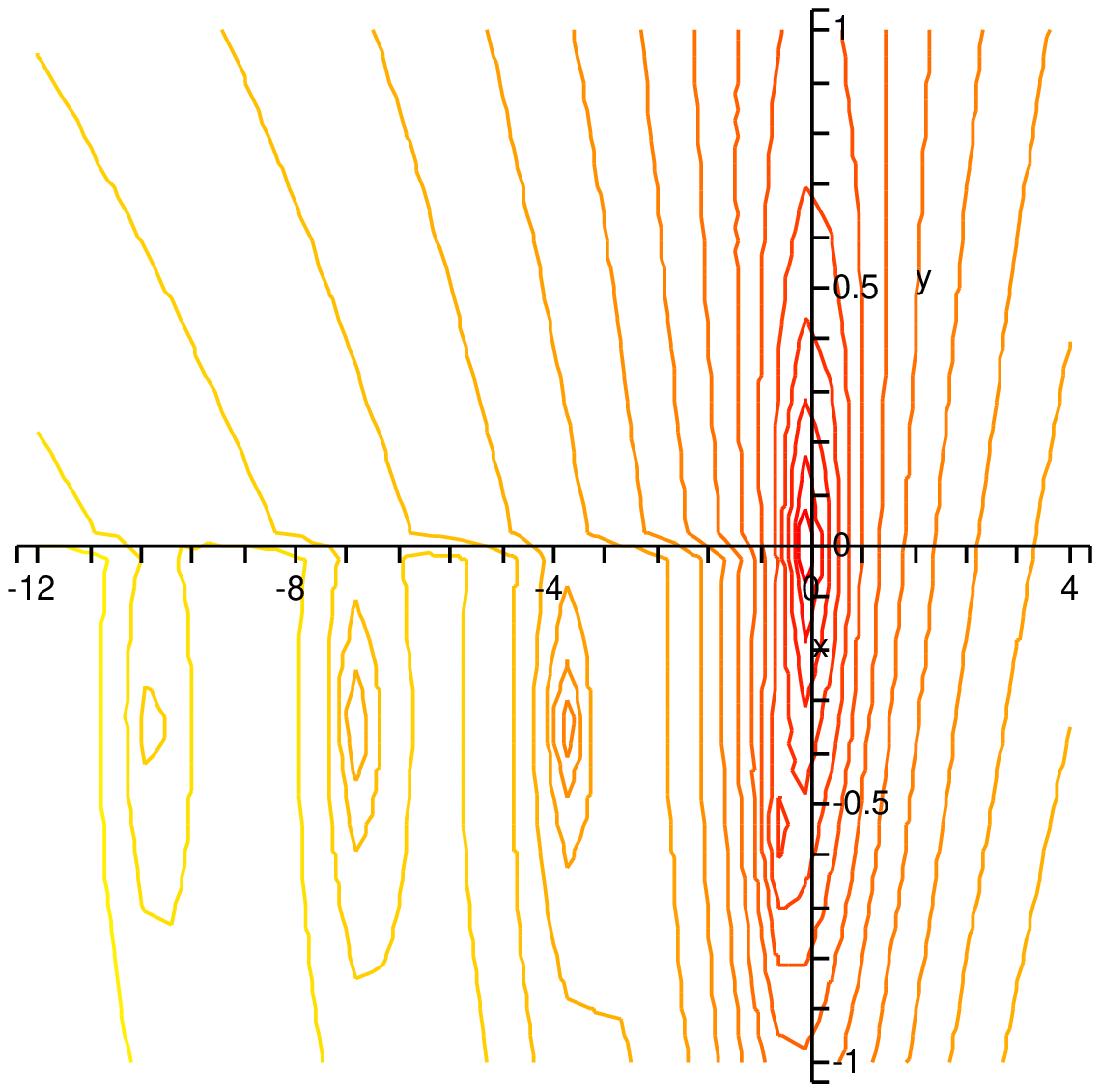}}
\end{center}
\end{minipage}
\begin{minipage}[t]{8cm}
\begin{center}
\resizebox{8cm}{8cm}{\includegraphics{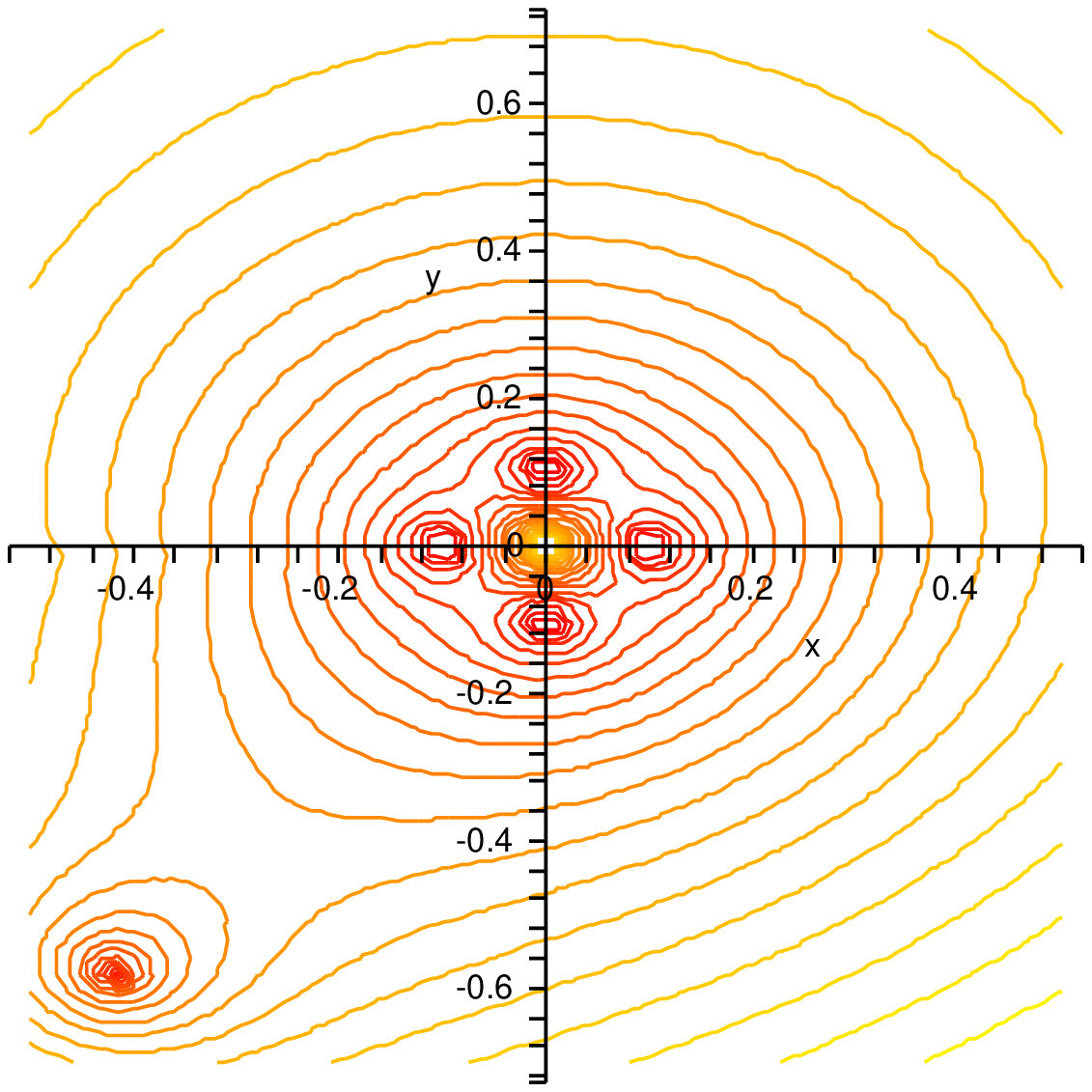}}
\end{center}
\end{minipage}
\caption{These plots, generated by the Maple 9.5 software package,
  show as poles the roots of the consistency relation when $\mu=m=0$
  and $\beta=0.01$.  They are contour plots of the logarithm of the
  absolute value of the l.h.s.~of Eq.~(\ref{consistency}).  The roots
  near the origin can be determined by Maple's root finding routine as
  $\pm (0.0997-1.40\times10^{-4}i)$ and $\pm 0.100i$.}
\label{Plots}
\end{figure}

%
The consistency relation (\ref{consistency}) can be solved numerically
for any given values of the parameters $m$, $\mu$ and $\beta$,
normalized by setting $\ell=1$.  The roots for massless fields and
small coupling, $\beta=0.01$, can be seen on in Fig.~\ref{Plots}.
The right hand plot shows the roots near the origin already calculated
analytically.  There are also other roots, one at $-0.419-0.577i$ and
a sequence with almost identical imaginary parts of $-0.355i$ exactly
analogous to those found by Seahra~\cite{Seahra}.

\subsection{Massive brane field}
Let us now consider the case were $\mu\gg\beta^2$ where we are still
considering $\beta\ll1$.  When $\beta=0$ there are roots to
Eq.~(\ref{consistency}) at $\r=\pm\mu$.  When $\beta\ne0$ we can work
out how these are perturbed.  By writing $\r=\mu+\delta\r$ we can
expand the consistency relation by a Taylor series about $\r=\mu$ to
get the lowest order correction
\begin{equation}
\delta\r = \frac{\beta^2\,H_\nu^{(1)}(\mu)}{4\mu\Big(\mu H_{\nu-1}^{(1)}(\mu)
+(2-\nu)H_\nu^{(1)}(\mu)\Big)}.
\end{equation}
So the roots are
\begin{equation}
\r\approx \pm\left(\mu+\frac{\beta^2\,H_\nu^{(1)}(\mu)}{4\mu\Big(\mu
H_{\nu-1}^{(1)}(\mu)
+(2-\nu)H_\nu^{(1)}(\mu)\Big)}\right),
\end{equation}
and the frequency $\omega$ will be given by the dispersion
relation~(\ref{omegar}). 

One case of special interest is when $m=0$, where this simplifies to
\begin{equation}
\r\approx
\pm\left(\mu+\frac{\beta^2}{4\mu^2}\frac{H_2^{(1)}(\mu)}{H_1^{(1)}(\mu)}\right)\,.
\end{equation}
The imaginary part always has the opposite sign to the real part.
We can use the well known asymptotic behaviour of Bessel
functions~\cite{GradRyz} to 
write
\begin{equation}
\Im \left( \frac{H_2^{(1)}(\mu)}{H_1^{(1)}(\mu)} \right)
\sim \left\{
\begin{array}{l@{\qquad}l}
-\frac{\pi}{2}\mu & \mu\to0\\
-1 & \mu\to\infty
\end{array}
\right.
\end{equation}
so for a light field on the brane, where $\beta\ll\mu\ll1$
\begin{equation}
\r\approx \pm\left(\mu-\frac{\pi\beta^2}{8\mu}\,i\right)
\,,\qquad
\omega \approx \pm \sqrt{k^2+\mu^2} -
\frac{\pi\beta^2}{8\sqrt{k^2+\mu^2}}\,i \,.
\end{equation}
There is also a solution with purely imaginary $\r$ which
corresponds to the tachyonic bound state.

\section{Conclusions}
\label{sect:conc}

In this paper we have investigated the behaviour of a boundary field
theory on a Minkowski brane linearly coupled to a bulk field in
anti-de Sitter spacetime. We have calculated an instability criterion
for the coupled boundary and bulk oscillators.  Above a critical
coupling $\beta_{\rm crit} = \sqrt{2(\nu-2)\mu^2/\ell}$ there is a
state with a purely imaginary effective mass.  This state is spatially
bounded and normalizable in the bulk, and describes a tachyonic
instability on large-scales (small $k$).  When either field is
massless $\beta_{\rm crit} = 0$, so that the instability will always
occur for non-zero coupling.  Below the critical coupling there is no
tachyonic state and no bound state, in contrast to the case of a
Minkowski bulk~\cite{George}. Instead we find quasi-normal modes.

The quasi-normal modes describe the late-time behaviour of the system
when there is no tachyonic instability.  
For weak coupling, there is a quasi-normal mode which represents a very slowly
decaying oscillatory solution. For instance in the low energy limit
($m\ell\ll1$) we find a decay time of the order $(m\ell)^{-1}m^{-1}$.
This contradicts the conjecture of Cartier and Durrer~\cite{CD04} that
orbifold boundary theories necessarily have instabilities.
It is necessary to impose a physical boundary condition in order to
determine which poles in the complex frequency plane describe the
behaviour of the system. Quasi-normal modes are determined by a
purely outgoing condition at infinity and thus only poles in the
lower half complex plane describe the late-time evolution of the system.

Quasi-normal modes are not normalizable modes.
For the classical problem normalizability is not a concern because the
quasi-normal mode solution is only valid in the causal future
development of the brane (see Ref.~\cite{Szpak} for a dicussion).
On the other hand normalisability is a concern if one wishes to
construct an initial quantum vacuum in terms of independent
oscillators with finite action.
In contrast to the work of George in a Minkowski bulk \cite{George} we
are not able to give the full spectrum of the quantum vacuum state
below the critical coupling. 

When the coupling is zero, a massless bulk field behaves in a way very
similar to gravitational waves in the Randall--Sundrum
brane-world~\cite{RS99}.  There is a massless zero mode solution on
the brane and a spectrum of massive Kaluza--Klein states. In addition,
there is a spectrum of quasinormal modes found by Seahra~\cite{Seahra}
which describe the evolution of the gravity waves at intermediate
times on the brane, though the zero mode dominates at late times. When
a non-zero coupling is introduced this behaviour is modified
dramatically.  The zero mode, $\r=0$, is no longer a solution but is
perturbed to solutions with complex $\r$.
Without the coupling, the boundary condition at the brane is purely
Neumann, thereby admitting a bounded zero-mode.  One can think of the
coupling as mixing Dirichlet and Neumann boundary conditions. A
similar effect could occur with a moving boundary, such as an expanding
brane universe.


An important difference in the case of gravitational perturbations in
the bulk is that these are not in general coupled to linear matter
perturbations on the brane and metric backreaction is treated as a
second-order effect. An important exception is the case of
fluctuations in an inflaton field driving inflation on a brane.
Inflaton fluctuations on the brane are linearly coupled to bulk metric
perturbations. These fluctuations start on small scales where they are
strongly coupled to bulk gravitational perturbations \cite{KLMW}.
Thus, we must specify the vacuum state for a coupled boundary (inflaton
fluctuations) and bulk (metric perturbations) system \cite{YK}.  

In order to address this issue, we need to extend our analysis to the
case of a de Sitter brane. Due to the curvature of a de Sitter brane,
there appears a mass gap \cite{GS} in the spectrum of the free bulk
field and there arises the possibility of having stable massive bound
states in the mass gap \cite{LS}. One could crudely model the
existence of a cosmological horizon in de Sitter space by imposing an
infra-red cut-off on wavelengths larger than the Hubble length,
suppressing the tachyonic instability for sufficiently small coupling
or large Hubble rate. We leave a full analysis of the de Sitter brane
for future work but it is intriguing to note that a de Sitter brane
might provide a good model for a coupled brane-bulk quantum field
theory, where a Minkowski brane does not.

\section*{Acknowledgments}

We would like to acknowledge very useful discussions with Sanjeev
Seahra, David Langlois, Misao Sasaki and Jiro Soda.
KK is supported by PPARC and AM by PPARC grant PPA/G/S/2002/00576.



\end{document}